\documentclass[prl,twocolumn,showpacs,preprintnumbers,amsmath,amssymb]{revtex4}
\usepackage{graphicx}
\usepackage{dcolumn}
\usepackage{bm}
\usepackage{psfig}
\newcommand \be{\begin{eqnarray}}
\newcommand \ee{\end{eqnarray}}
\begin{document}
\title{Discontinuity of capacitance at the onset of surface superconductivity} 
\author{K. Morawetz$^{1,2}$, P. Lipavsk\'y$^{3,4}$ and J. J. Mare{\v s}$^{4}$}
\affiliation{$^1$Forschungszentrum Dresden-Rossendorf, PF 51 01 19, 
01314 Dresden, 
Germany}
\affiliation{
$^2$Max-Planck-Institute for the Physics of Complex
Systems, Noethnitzer Str. 38, 01187 Dresden, Germany}
\affiliation{
$^3$Faculty of Mathematics and Physics, Charles University, 
Ke Karlovu 3, 12116 Prague 2, Czech Republic}
\affiliation{
$^4$Institute of Physics, Academy of Sciences, 
Cukrovarnick\'a 10, 16253 Prague 6, Czech Republic
}
\begin{abstract}
The effect of the magnetic field on a capacitor with a superconducting 
electrode is studied within the Ginzburg-Landau approach. It is shown 
that the capacitance has a discontinuity at the onset of the surface 
superconductivity $B_{\rm c3}$ which is expressed as a discontinuity in the 
penetration depth of the electric field into metals. Estimates show that 
this discontinuity is observable with recent bridges for both conventional 
and high-$T_{\rm c}$ superconductors of the type-II.
\end{abstract}
\maketitle

Capacitors based on ferroelectric layers sandwiched between metallic 
electrodes are approaching the technical limits of their performance.
Their capacitance is not anymore exclusively given by
the dielectric response of the isolating ferroelectric layer but it
is reduced due to the penetration of the electric field into the 
metallic electrodes. Concerning the large scale integration 
of microscopic capacitors the penetration of the electrostatic field 
into the electrodes is considered as a lumped series capacitance. 
From the viewpoint of fundamental research, however, this phenomenon offers an 
opportunity to study the interaction of metallic surfaces with an applied 
electric field. 

Here we discuss a possibility to observe the penetration 
of the electrostatic field into the metal in the vicinity of the 
transition from the normal to the superconducting state. We focus on
the third critical magnetic field $B_{\rm c3}$ 
at which field the superconducting state nucleates 
at the surface. We predict that at this field the 
capacitance or the penetration of the electrostatic 
field possesses a jump.

The penetration of the electrostatic field into the normal metal is
well understood. Ku and Ullman \cite{KU64} have derived an analytic 
solution of the penetrating field for the jellium model within the 
Thomas-Fermi approximation. Their simple prediction is sufficient 
to explain the experimental data \cite{BW99}. Much less is known about 
the penetration of the electrostatic field into
superconductors. From the very beginning until now the history of this 
problem is full of contradictory concepts yielding a wide scattering
of predicted values. 

The question of the penetration of the electrostatic field into 
superconductors has been firstly addressed by the Lon\-don brothers. 
In their early paper in 1935 they have expected that the penetration depths 
of the electrostatic and magnetic fields are identical \cite{L35}. One year 
later H. London measured the capacitance with superconducting electrodes 
controlled by  the magnetic field and concluded that the penetration 
of the electrostatic field into the metal is not changed by the 
transition to the superconducting state \cite{HL36}. While the former 
concept predicts thousands of \AA ngstr\"oms for conventional 
superconductors, the latter concept suggests less than one \AA ngstr\"om.

Oppositly, the way from small to large penetration depths 
one meets in the papers by Anderson
and coworkers. The Anderson theorem \cite{Anderson59} 
states that the thermodynamical properties of the superconducting 
condensate do not depend on the electrostatic field. Accordingly, 
the condensate does not affect the penetration of the electrostatic 
field which is thus the same as in the normal metal. More recently, 
in the brief discussion of the effect observed by Tao, Zhang, 
Tang and Anderson \cite{Tao99}, the authors speculate about a~large
penetration depth of the electrostatic field using ideas of the
Anderson model \cite{Anderson98} of the high-$T_{\rm c}$ 
superconductivity.

Apparently the problem of the electrostatic field penetrating the
surface of the superconductor is far from being settled and a clear
experimental message is still missing. In this letter we propose 
an experiment on the ferroelectric capacitor with one normal and
one superconducting electrode. The magnetic field is applied to
switch off the superconductivity and we will explore the vicinity
of the third critical field $B_{\rm c3}$.

The sensitivity of ferrroelectric devices to the screening in metals is
striking. Indeed, the typical Thomas-Fermi screening length in metals
is about $0.5\,$\AA, while the width $L$ of the insulating layer has 
to be about a thousand of \AA ngstr\"oms to guarantee low leakage 
currents. The direct comparison of these scales is somewhat misleading, 
however. Taking into account the dielectric constants of the components
involved we immediately obtain
\be
{\delta C \over C}={\epsilon_{\rm d}\over \epsilon_{\rm s}} 
{\delta L \over L}.
\label{e0}
\ee  
The ceramic ferroelectric materials have $\epsilon_{\rm d}
\sim 10^3$ and metals have the
ionic background permittivity $\epsilon_{\rm s}\sim 4$ giving
an enhancement factor $\epsilon_{\rm d}/\epsilon_{\rm s}\sim 250$.
The capacitance can be measured with sensitivity better than 
$\delta C/C\sim 10^{-6}$, which makes it possible to observe very
subtle changes of the penetration depth $\delta L\sim 10^{-5}$\AA .

First let us take a look at the interaction between the electrostatic 
field and the superconductivity. The superconducting surface under 
the applied electrostatic field has been theoretically studied at 
various levels by Nabu\-tovsky and Shapiro 
\cite{NS81,NS82,Shapiro84,Shapiro85}. They have shown that 
the phenomenological theory of Ginzburg and Landau (GL) yields 
basically the same result as the microscopic picture based on the 
Bogoliubov-de Gennes method. Their result was recovered in a simple form
in Ref.~\cite{LMKY06}, where it was shown that the effect of the applied 
electrostatic field $E$ merely modifies the extrapolation length $b$
in the de Gennes boundary 
condition for the GL function $\psi$,
\begin{align}
{\nabla\psi\over\psi}={1\over b}={1\over b_0}+{E\over U_{\rm s}}.
\label{e1}
\end{align}
This field effect is measured on the voltage scale
\begin{align}
{1\over U_{\rm s}}=
\kappa^2{\partial\ln T_{\rm c}\over\partial\ln n}
{e^*\epsilon_{\rm s}\over m^*c^2},
\label{e2}
\end{align}
where $\kappa$ is the GL parameter. The logarithmic derivative of
the critical temperature with respect to the electron density is of the order
of unity. The need for strong applied fields follows from the 
`relativistic' energy of an electron which is rather large,
$m^*c^2\sim 1$~MeV.

The boundary condition (\ref{e1}) restricts the solution of the
GL equation
\begin{align}
{1\over 2m^*}\left(-i\hbar\nabla-e^*{\bf A}\right)^2\psi+
\alpha\psi+\beta|\psi|^2\psi=0
\label{e3}
\end{align}
at the surface. Being non-linear, the GL equation 
(\ref{e3}) has the ability to heal any perturbation of the GL 
function from its optimal value on the GL coherence length 
$\xi=\hbar/\sqrt{2|\alpha| m^*}$. The boundary 
condition thus affects the GL function only in the vicinity of the surface.
As consequence of that, the electric field has no remarkable effect on the 
bulk superconductivity.

Saint-James and de Gennes \cite{SJdG63} have noticed that simi\-larly 
to the condensation of the vapor at surfaces, the boundary 
condition (\ref{e1}) implies a nucleation of the superconducting 
condensate at the surface. This becomes apparent at high magnetic
fields, since the bulk superconductivity vanishes at the upper
critical field $B_{\rm c2}$ while a thin sheet of superconducting
condensate survives near the surface up to fields $B_{\rm c3}
\sim 1.69461\,B_{\rm c2}$. Their result applies to the infinite 
extrapolation length, $1/b_0+E/U_{\rm s}=0$.

Let us modify the method of Saint-James and de Gennes for a finite
$b$ or non-zero $E$. For the third critical field the GL function
has an infinitesimally small amplitude so that one can neglect the
cubic term in (\ref{e3}). As the diamagnetic current is also negligible,
the vector potential reads
${\bf A}=(0,B_{\rm c3}x,0)$. We have associated the surface with 
the plane $x=0$. The GL equation (\ref{e3}) is then solved by the 
parabolic cylinder function of Whittaker \cite{MOS66}
\begin{align}
\psi(x,y,z)=N\, {\rm e}^{iky}
D_{\nu-{1\over 2}}\left({2x\over l}-kl\right),
\label{e4}
\end{align}
with $x$ scaled by the magnetic length $l^2=\hbar/(eB_{\rm c3})$ 
and $\nu=-\alpha m^*/\hbar e^*B_{\rm c3}= 
{l^2}/{(2\xi)^2}$. 

\begin{figure}
\psfig{file=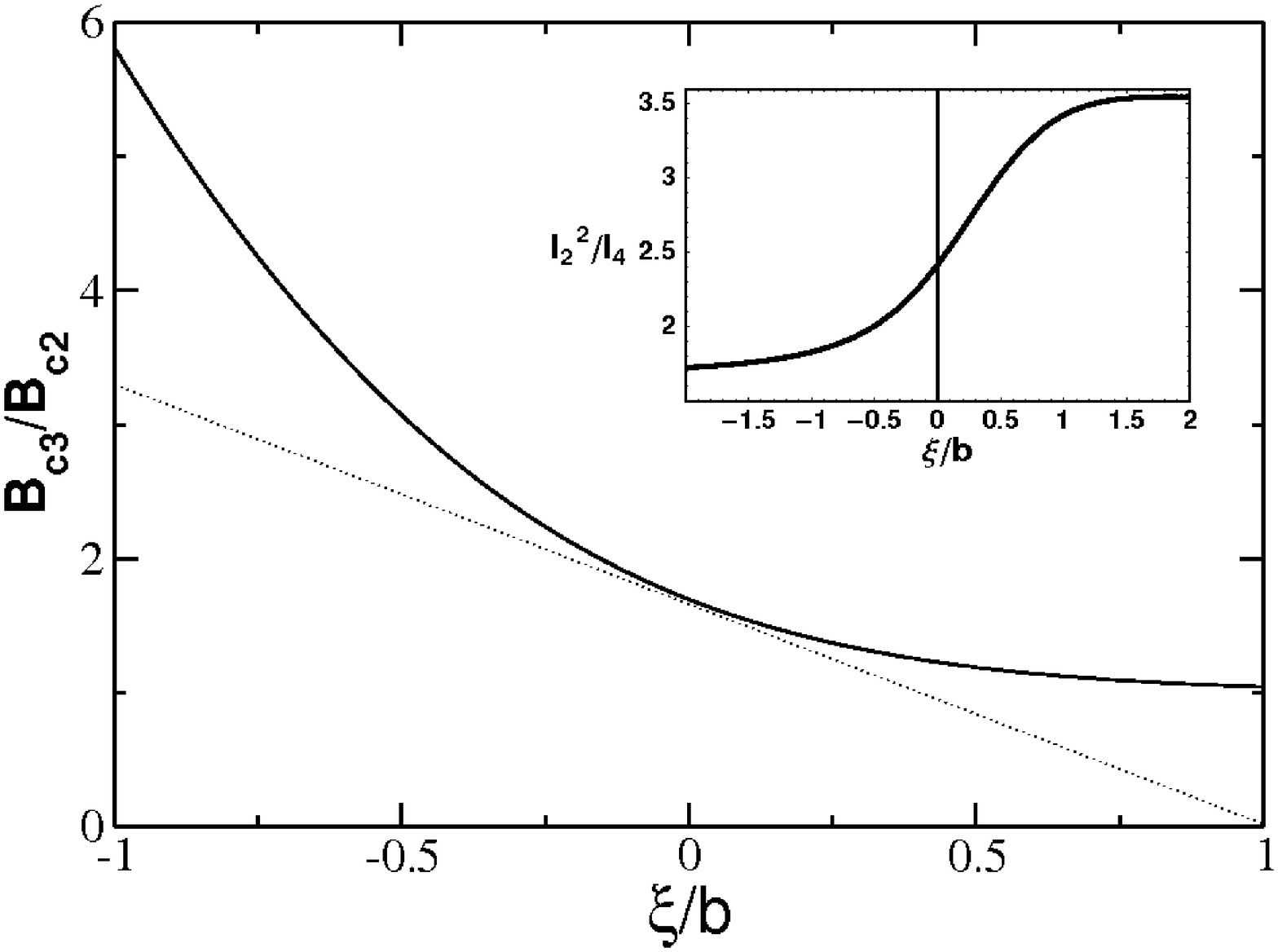,width=9cm}
\vspace*{-3ex}
\caption{The surface critical field $B_{\rm c3}$ versus the
applied electric field $E$ (solid line). The ratio of the third to the upper
critical magnetic field relates to the argument of the parabolic
cylinder function, $B_{\rm c3}/B_{\rm c2}=2\nu$. The tangential line at zero bias is 
given as dotted line. The inset shows the field dependence of the factor in (\ref{e9}).
}
\label{bc3f}
\end{figure}

So far $\nu$ and $k$ are parameters of the nucleating GL 
function $\psi$. We are looking for the solution with the lowest 
$\nu=\nu_{\rm min}$, because it corresponds to the highest magnetic field 
$B_{\rm c3}$ for which the nucleation is possible at fixed 
temperature. This highest magnetic field $B_{\rm c3}$ is the third 
critical field and its resulting value is shown in figure~\ref{bc3f}. 

The dependence of the third critical field $B_{\rm c3}$ on the
applied electrostatic field $E$ indicates that the electrostatic field 
affects the surface superconductivity. The same interaction
manifests itself in the effect of the magnetic field on the
capacitance. 
The capacitance of the capacitor with one superconducting electrode 
reads
\begin{align}
{1\over C_{\rm s}}={1\over C_{\rm n}}-
{1\over\epsilon_0^2\epsilon_{\rm s}^2S^2}
{\partial^2 F\over\partial E^2},
\label{e5}
\end{align}
where $S$ is the area of the capacitor, $C_{\rm n}$ is the 
capacitance when both electrodes are normal, and 
\begin{align}
F=S\!\int\limits_0^\infty \! dx
\left[\,{1\over 2m^*}\left|
\left(i\hbar\nabla\!+\!e^*{\bf A}\right)\psi\right|^2\!+\!
\alpha|\psi|^2\!+\!{\beta\over 2}|\psi|^4\right]
\label{e6}
\end{align}
is the GL free energy describing the difference between
the normal and the superconducting states in the superconducting electrode.

Near the transition line $B\sim B_{\rm c3}$, the GL function has the 
shape of the nucleation function (\ref{e4}) with $\nu=\nu_{\rm min}(E,B)$ 
and $k=k_{\rm min}(E,B)$. If we keep the amplitude $N$ as a variational 
parameter, the free energy is its biquadratic function, 
$F_N=S(\alpha-\alpha_E)lN^2I_2/2+SlN^4I_4/(4\beta)$. Here 
$\alpha_E=-(\hbar e^*B/m^*)\nu_{\rm min}(E,B)$ stands for the
kinetic energy obtained from the GL function (\ref{e4}), while 
$\alpha=\alpha'(T-T_{\rm c})$ is the temperature dependent GL
parameter. Integrals 
over powers of the parabolic
cylinder functions are denoted as $I_n=\int_{k_E}^\infty d\tau\,
D_{\nu_E}^n$. The condition of minimum, $\partial F_N/\partial 
N=0$, is satisfied by $N_{\rm min}^2=(\alpha_E-\alpha)I_2/(\beta I_4)$.
From $F=F_{N_{\rm min}}$ or directly from (\ref{e6}) one obtains the
free energy $F=-S(\alpha_E-\alpha)^2lI_2^2/(4\beta I_4)$.

Now we can evaluate the jump of the capacitance, which appears 
as the magnetic field $B$ exceeds the critical value $B_{\rm c3}$. 
Since $\alpha_E\to\alpha$ for $B\to B_{\rm c3}$, the discontinuity of
the inverse capacitance equals
\begin{align}
{1\over C_{\rm s}}-{1\over C_{\rm n}}=
{\hbar^2 e^{*2}B_{\rm c3}^2 l\,I_2^2\over
2\epsilon_0^2\epsilon_{\rm s}^2m^{*2}S\beta I_4}
\left({\partial\nu_{\rm min}\over\partial E}\right)^2,
\label{e7}
\end{align}
where we have used ${\partial\alpha_E/\partial E}=-(\hbar e^*B/m^*)
(\partial\nu_{\rm min}/\partial E)$. The tangential line plotted in Fig.~\ref{bc3f} yields $\partial\nu_{\rm min}/\partial E=-
0.82\,\xi/U_{\rm s}$. 
The discontinuity in the capacitance is transparently expressed via
the discontinuity in the penetration depth of the electric field
\begin{align}
\delta L=\epsilon_0\epsilon_{\rm s}S
\left({1\over C_{\rm s}}-{1\over C_{\rm n}}\right)=
{0.397 \,\hbar^4\,I_2^2\over 
\epsilon_0\epsilon_{\rm s}m^{*2}\beta U_{\rm s}^2l\,I_4}.
\label{e9}
\end{align}
In the rearrangement we have used $4\xi^2=l^2/\nu$ and
$e^*B_{\rm c3}=2 \hbar/l^2$. From $2\nu=1.694$ follows the
numerical factor $0.82^2/(2\nu)=0.397$. The field dependence of the 
factor $I_2^2/I_4$ is plotted as an inset in figure~\ref{bc3f}.

To further simplify expression (\ref{e9}) we employ the 
parameter $\beta=6\pi^2k_{\rm B}^2T_{\rm c}^2/(7\zeta(3)E_{\rm F}n)$
derived form the BCS theory by Gor'kov \cite{Gor59}, and rewrite it
in terms of the \mbox{BCS} coherence length $\xi_{\rm BCS}=\hbar v_{\rm F}/
(1.76 \pi \,k_{\rm B}T_{\rm c})$. Moreover we substitute $U_{\rm s}$ 
from (\ref{e2}) so that we obtain finally
\begin{align}
\delta L=1.86\,10^{-8}{\kappa^4\epsilon_{\rm s}
\over m_{\rm s}^3}a_{\rm B}^3n
\left({\partial\ln T_{\rm c}\over\partial\ln n}\right)^2
{\xi_{\rm BCS}^2\over l}.
\label{e10}
\end{align}
We have collected all universal physical constants into the Bohr 
radius $a_{\rm B}=4\pi\hbar^2\epsilon_0/(m_0e^2)=0.53\,$\AA\ and 
the constant of fine structure $e^2/(4\pi\epsilon_0\hbar c)=1/137$. 
The later appears in the fourth power giving the very small factors 
$1/137^4=2.8\times 10^{-9}$. The mass of the Cooper pair is twice the 
effective mass of electrons in the metal $m^*=2m_{\rm s}m_0$ and
$e^*=2e$. For $1/b_0+E/U_{\rm s}=0$ the factor given by the profile 
of the GL function is $I_2^2/I_4=2.42$, see figure~\ref{bc3f}. 

Equation 
(\ref{e10}) is the main result of the paper. It expresses the jump in the 
capacitance (\ref{e9}) in terms of materials parameters like the logarithmic 
density derivative of the critical temperature, the coherence length 
$\xi_{\rm BCS}$ and the GL parameter $\kappa$. This result provides a 
convenient tool to access these parameters by measuring the jump in the 
capacitance at the third critical field $B_{\rm c3}$.
Indeed, the discontinuity is small but nevertheless observable. 

For an 
estimate we assume some typical numbers. The most sensitive measurements of 
capacitance performed in the $C\sim\mu$F range are capable to monitor
changes $\delta C/C\sim 10^{-6}$ with error bars at  $\delta C/C
\sim 10^{-7}$. From the capacitance $C=\epsilon_0\epsilon_{\rm d}S/L$ 
one sees that a 1000\,\AA-thick dielectric layer with 
$\epsilon_{\rm d}=10^3$ has an optimal area of 10~mm$^2$ which 
is about the usual size of such samples \cite{Hwang02}.
The penetration depth (\ref{e9}) yields the relative change of the
capacitance according to (\ref{e0}). With $\epsilon_{\rm s}=4$ and the 
above assumed values for the capacitance one finds that changes 
$|\delta L|>3\times10^{-6}$\AA\ are conveniently detectable with error bars
of $\delta L\sim 3\times10^{-7}$\AA.

It should be noted here that these estimates remain essentially valid
even if the Wagner polarization diminishing effective permittivity of
thin dielectric layers is taken into account. The expected reduction 
of the numbers above corresponds only to a factor of $\sim 2$,
see Ref.~\cite{NOS98}.

Now we will show that for niobium the discontinuity falls 
in the range of the error bars. For niobium at temperature
$T\sim 1$\,K one can take $\kappa\sim 1.5$, see \cite{FSS66}, and 
$m_{\rm s}=1.2$ giving $\kappa^4\epsilon_{\rm s}/m_{\rm s}^3=11.7$. 
The logarithmic derivative is estimated in \cite{LKMBY07} as
${\partial\ln T_{\rm c}/\partial\ln n}=0.74$. The electron density
$n=2.2\times 10^{28}$/m$^3$ yields $a_{\rm B}^3n=3.3\times 10^{-3}$ and the Fermi 
velocity $v_{\rm F}=\hbar(3\pi^2n)^{1/3}/(m_0m_{\rm s})=7.2\times 10^5$~m/s. 
The critical temperature $T_{\rm c}=9.5$\,K corresponds to the \mbox{BCS}
coherence length of $\xi_{\rm BCS}=3120$\,\AA. Finally we need the third 
critical magnetic field $B_{\rm c3}$ to estimate the magnetic
length $l$. From $B_{\rm c3}=1.69\,B_{\rm c2}$ and the experimental value 
$B_{\rm c2}=0.35$\,T \cite{FSS66} one finds $B_{\rm c3}=0.59$\,T,
which yields $l=325$\,\AA. With all these values we obtain from
equation (\ref{e10}) the discontinuity 
$\delta L\sim 1.2\times 10^{-6}$\AA, which is comparable to the error bar.

There are a number of alloys \cite{BH63} with the help of which one easily 
reaches a region of observable discontinuities. For example, 50\% of niobium 
with 50\% of tantalum has the critical temperature $T_{\rm c}=6.25$\,K 
while the GL parameter at $T_{\rm c}$ is $\kappa=3.9$ \cite{MS65}. 
The upper critical magnetic field at $T\ll T_{\rm c}$ is $0.7$\,T 
\cite{MS65} giving $B_{\rm c3}=1.2$\,T which yields $l=228$\,\AA. 
We assume that the effective mass scales with the GL parameter 
so that $\kappa/m_{\rm s}$ remains the 
same as in pure niobium along with the remaining parameters. 
In this case, the discontinuity increases 
to $\delta L= 1.1\times 10^{-5}$\AA, which is still well 
observable.

We note that among intermetallic alloys there are even more promising
candidates. The alloy of Nb-61\% Ti has $T_{\rm c}=8.95$\,K and 
$\kappa=38.4$. The upper critical magnetic field $B_{\rm c2}=47$\,T
corresponds to $B_{\rm c3}=80$\,T, which is too high to be applied
during slow measurements of the capacitance. One has to increase 
the temperature for the measurement so that the third critical field 
becomes comparable to a convenient field of 10~T, which corresponds 
to $l=79$\,\AA. This estimate suggests $\delta L=1.5\times 10^{-4}$\AA\
which is fifty times larger than the experimental sensitivity.

Detectable amplitudes of the discontinuity result also for 
high-$T_{\rm c}$ materials. Using the values of YBa$_2$Cu$_3$O$_{7-\delta}$ 
which are
$T_{\rm c}=90$\,K, $\kappa=55$, $m_{\rm s}=6.92$, $\epsilon_{\rm s}=4$, 
$n=5\times 10^{27}$/m$^3$ \cite{Plak95} and ${\partial\ln T_{\rm c}/
\partial\ln n}=-2.4$ \cite{lnTclnn} as well as $l=79$\,\AA\ for $10$\,T of 
the applied magnetic field, one can expect a discontinuity 
$\delta L=1.4\times 10^{-5}$\AA. 

It should be noted
that the field effect on the high-$T_{\rm c}$ materials has been 
extensively studied within the effort to develop superconducting 
devices analogous to the field-effect transistors \cite{FMBW95,ATM03}. 
There were many measurements of the field effect detecting directly
changes in $T_{\rm c}$ with the applied electric field. These 
experiments employ the largest accessible fields because the changes in
$T_{\rm c}$ are very small. The discontinuity of the capacitance 
can supply the missing knowledge of the field effect for low applied 
fields. Since the mechanism of the field effect on the high-$T_{\rm c}$ 
materials is not yet fully clarified, the low-field effect is of 
interest.

In summary, we have shown that the capacitance of the planar 
capacitor with one normal electrode and the other electrode 
to be superconducting
possesses a discontinuity at the third critical field $B_{\rm c3}$. 
This discontinuity is large enough to be observed in capacitors 
with ferroelectric dielectric layers of a width of 1000\,\AA. 
We would like to point out that compared to other regions of the 
magneto-capacitance, the discontinuity has the advantage of being 
a unique feature which is not obscured by other properties of the insulator. 
Indeed, exploring strong electric fields one has to face the fact that the 
dielectric response of the ferroelectric material is non-linear. 
Scanning through temperatures one observes namely the Curie law of 
the ferroelectric transition. Moreover, the dielectric function of 
the ferroelectric isolator depends on the magnetic field. The 
measurement of the discontinuity circumvents all these problems, 
because all the troublesome dependencies are continuous at 
the onset of the surface superconductivity.

\medskip
This work was supported by research plans MSM 0021620834 and 
No. AVOZ10100521, by grants GA\v{C}R 202/07/0597 and 202/06/0040 
and GAAV 100100712 and IAA1010404, by PPP project of DAAD, by DFG 
Priority Program 1157 via GE1202/06 and the BMBF and by European 
ESF program NES.

\bibliography{kmsr,kmsr1,kmsr2,kmsr3,kmsr4,kmsr5,kmsr6,kmsr7,delay2,spin,gdr,refer,sem1,sem2,sem3,micha,genn,solid,deform}

\end{document}